\documentclass[a4paper]{PoS}
\usepackage{amssymb}
\usepackage{amsmath}
\usepackage{url}

\usepackage{graphicx}
\graphicspath{{./Figure/}}

\def\units#1{~\hbox{$\,{\rm #1}$}}

\title{Hadronic interactions of primary cosmic rays with the \texttt{FLUKA} code}

\ShortTitle{Hadronic interactions of CRs with \texttt{FLUKA}}

\author{\speaker{M.~N.~Mazziotta} \\
        Istituto Nazionale di Fisica Nucleare, Sezione di Bari, 70126 Bari, Italy \\
        E-mail: \email{mazziotta@ba.infn.it}}

\author{F.~Cerutti\\
        CERN, Geneva, Switzerland \\
	E-mail: \email{Francesco.Cerutti@cern.ch} }

\author{A.~Ferrari \\
        CERN, Geneva, Switzerland \\
	E-mail: \email{Alfredo.Ferrari@cern.ch} }

\author{D.~Gaggero \\
        SISSA, via Bonomea 265, 34136 Trieste, Italy \\
	INFN, Sezione di Trieste, via Valerio 2, 34127 Trieste, Italy \\
	E-mail: \email{Daniele.Gaggero@sissa.it} }

\author{F.~Loparco \\
        Dipartimento di Fisica ``M. Merlin" dell'Universit\`a e del Politecnico di Bari, I-70126 Bari, Italy \\
	Istituto Nazionale di Fisica Nucleare, Sezione di Bari, 70126 Bari, Italy \\
	E-mail: \email{Francesco.Loparco@ba.infn.it} }

\author{P.~R.~Sala \\
	Istituto Nazionale di Fisica Nucleare, Sezione di Milano, 20133 Milano, Italy \\
	E-mail: \email{Paola.Sala@mi.infn.it} }
		
\abstract{
The measured fluxes of secondary particles produced by the interactions 
of cosmic rays with the astronomical environment represent a powerful tool to infer 
some properties of primary cosmic rays. In this work we investigate the production of 
secondary particles in inelastic hadronic interactions between several 
cosmic rays species of projectiles and different target nuclei of the 
interstellar medium. The yields of secondary particles have been 
calculated with the \texttt{FLUKA} simulation package, that provides with very 
good accuracy the energy distributions of secondary products in a large energy range. 
An application to the propagation and production of secondaries in the 
Galaxy is presented.
}

\FullConference{The 34th International Cosmic Ray Conference,\\
		30 July- 6 August, 2015\\
		The Hague, The Netherlands}

\begin{document}

\section{Introduction}
According to the currently accepted scenario, Galactic Cosmic Rays (CRs) are accelerated in Supernova Remnants (SNRs) 
and interact with the turbulent interstellar magnetic field: the resulting motion is well described by a diffusion equation. 
During their journey through the Galaxy, the inelastic collisions of hadronic CRs with the interstellar medium (ISM), 
usually dubbed {\it spallation} processes, produce lighter particles and secondary radiation: the study of these events 
is very important since it may shed light on the origin of the CRs themselves and on the mechanisms governing their transport.

The production of light nuclei -- such as Boron -- from heavier ones (in particular Carbon and Nitrogen) has been 
extensively studied in recent times, since light nuclei ratios (e.g. B/C, N/O) are often used to constrain the propagation 
models in the Galaxy and in particular the rigidity dependence and normalization of the diffusion coefficient. 

The collisions of protons and Helium nuclei with the gas, and subsequent decays of the produced neutral pions ($\pi^0$s), 
are expected to give the most relevant contribution to the gamma-ray diffuse emission in the Galactic plane, since that is the 
region with the largest gas column densities. 
In this context, the high precision gamma-ray maps and spectra measured by the Fermi-LAT instrument in the energy band spanning 
from several tens of~\units{MeV} to several hundreds of~\units{GeV} are very useful and represent a unique opportunity to understand 
the CR transport properties in different regions of the Galaxy~\cite{FermiLAT:2012aa}. 

Charged leptons and antiprotons are another important product of the inelastic hadronic collisions
of CRs with the interstellar gas. The PAMELA collaboration 
measured the positron, electron and antiproton spectra, as well as the spectra of many light
nuclei~\cite{pamela}.
Recently the AMS02 collaboration provided very accurate results on the positron fraction and new 
results on the positron, electron and proton intensities, as well as on the helium spectrum and on the 
B/C ratio~\cite{ams}. All those datasets were taken during the same period, 
and this circumstance makes the interpretation of the data easier. Simultaneous 
measurements of several particle spectra performed by the same experiment over a wide energy 
range will in fact ensure reduced experimental systematics and will also limit the uncertainties 
arising from the CR propagation in the heliosphere (solar modulation) and in the Galaxy. 


A common practice in the literature is to fit different observables (e.g. hadronic and leptonic spectra) using
cross sections obtained from several parameterizations, derived with different methods and under different assumpions: this may
lead to inconsistencies in the determination of CR diffusion models.
In the present work we aim at providing a complete and consistent set of cross sections for the secondary production
of hadrons and leptons. 
We perform our study with the {\tt FLUKA} MC simulation code~\cite{fluka}, and we present a 
comprehensive calculation of the secondary hadron, lepton, gamma-ray and neutrino 
yields produced by the inelastic interactions between several species of stable or long-lived cosmic rays projectiles 
(p, D, T, $^{3}$He, $^{4}$He, $^{6}$Li, $^{7}$Li, $^{9}$Be, $^{10}$Be, $^{10}$B, $^{11}$B, $^{12}$C, $^{13}$C, 
$^{14}$C, $^{14}$N, $^{15}$N, $^{16}$O, $^{17}$O, $^{18}$O, $^{20}$Ne, $^{24}$Mg and $^{28}$Si)
and different target gas nuclei (p, $^{4}$He, $^{12}$C, $^{14}$N, $^{16}$O, $^{20}$Ne, $^{24}$Mg and $^{28}$Si).

The present results provide, for the first time, a complete and self-consistent set of all the relevant inclusive cross sections regarding 
the whole spectrum of secondary products in nuclear collisions. 
We cover, for the projectiles, a kinetic energy range extending from $0.1\units{GeV/n}$ up to $100\units{TeV/n}$ in the lab frame~\cite{mazziotta2015}. 
In order to show the importance of our results for multi-messenger studies about the physics of CR propagation, we evaluate the propagated spectra 
of Galactic secondary nuclei, leptons, and gamma rays produced by the interactions of CRs with the insterstellar gas, exploiting the numerical 
code \texttt{DRAGON}~\cite{dragon}.
 
\section{Secondary particle production in CR interactions with {\tt FLUKA}}
{\tt FLUKA} is a general purpose Monte Carlo (MC) code for the simulation of hadronic and electromagnetic interactions. 
It is used in many applications, and is continuously checked using the available data from low energy nuclear physics, high-energy 
accelerator experiments and measurements of particle fluxes in the atmosphere. 
Hadronic interactions are treated in {\tt FLUKA} following a theory-driven approach~\cite{ferrari1996}. 
The general phenomenology 
is obtained from a microscopical description of the interactions between the fundamental constituents 
(quarks and nucleons), appropriate for the different energy regions. 
Below a few $\units{GeV}$, the hadron-nucleon interaction model is based on resonance 
production and decay of particles, while for higher energies the Dual Parton Model (DPM) is used. 
The extension from hadron-nucleon to hadron-nucleus interactions is done in the framework of
the PreEquilibrium Approach to NUclear Thermalization model ({\tt PEANUT})~\cite{peanut},
including the Gribov-Glauber multi-collision mechanism followed by the pre-equilibrium 
stage and possible equilibrium processes (evaporation, fission, Fermi break-up and gamma deexcitation). 

\begin{figure*}[ht!]
\includegraphics[width=0.5\columnwidth,keepaspectratio,clip]{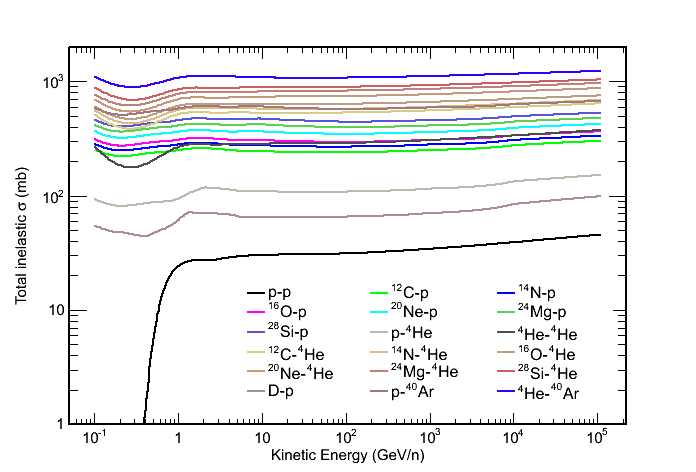}
\includegraphics[width=0.5\columnwidth,keepaspectratio,clip]{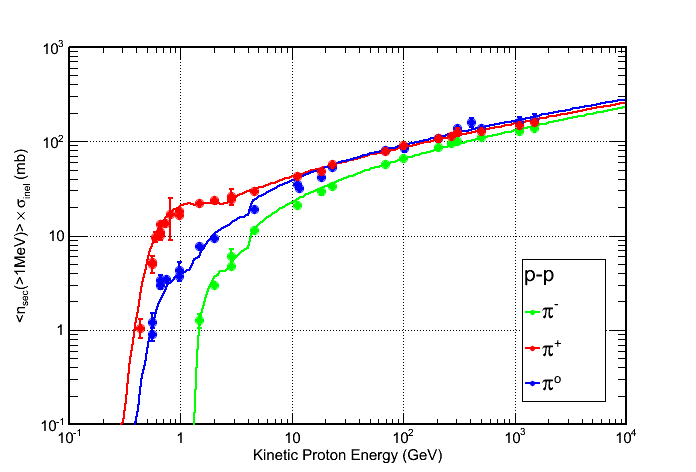}
\caption{Left panel: Total inelastic cross sections as a function of the energy per nucleon of the incoming projectile.
The plot shows the cross sections for all the projectile-target pairs studied in the present work.
Right panel: Inclusive cross sections for the production of $\pi^0$ (blue), $\pi^{+}$ (red) and $\pi^{-}$ (green)
in p-p collision as function of the incoming proton kinetic energy. 
Lines: {\tt FLUKA} simulation; points: data from Ref.~\cite{dermer1986}.}
\label{FigXsecInelPion}
\end{figure*}


The first quantities we want to investigate are the total inelastic cross sections for the collisions
we are focusing on, and -- as an important cross check -- the pion production cross section (for the proton-proton case)
compared to current data.
The left panel of Fig.~\ref{FigXsecInelPion} shows the total inelastic cross section calculated in the {\tt FLUKA} code for the
collision processes investigated in the present work.
The right panel of the Fig.~\ref{FigXsecInelPion} shows the pion inclusive cross sections in p-p collisions calculated from the {\tt FLUKA} 
simulation at the interaction level compared to current data.

%
%
\begin{figure*}[!ht]
\includegraphics[width=0.325\columnwidth,keepaspectratio,clip]{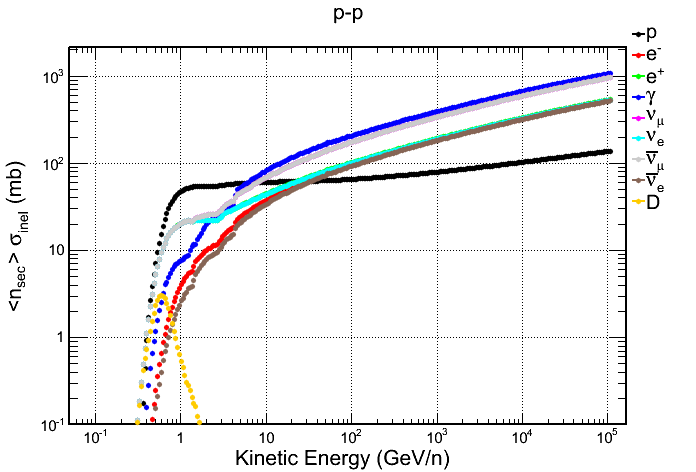} 
\includegraphics[width=0.325\columnwidth,keepaspectratio,clip]{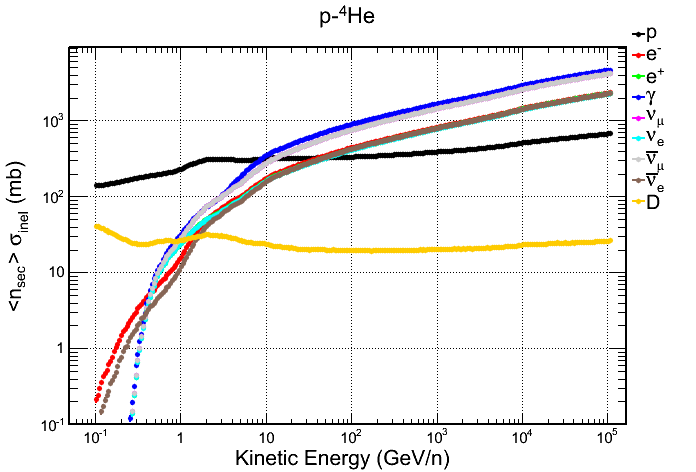}
\includegraphics[width=0.325\columnwidth,keepaspectratio,clip]{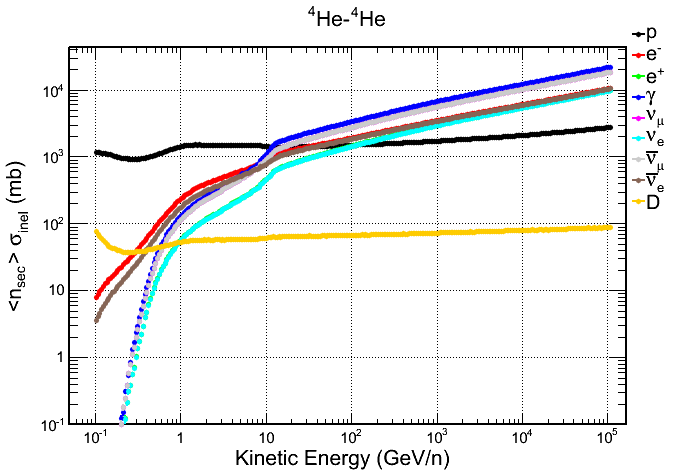}
\caption{Inclusive cross sections for the production of protons (black), electrons (red), 
positrons (green), gamma rays (blue), electron neutrinos (cyan), 
electron antineutrinos (grey), muon neutrinos (magenta), muon antineutrinos (brown)
and Deuterons (orange) in the p-p, p-$^{4}$He and $^{4}$He-$^{4}$He collisions.}
\label{FigpHeInclXsec}
\end{figure*}

\begin{figure*}[!ht]
\includegraphics[width=0.325\columnwidth,keepaspectratio,clip]{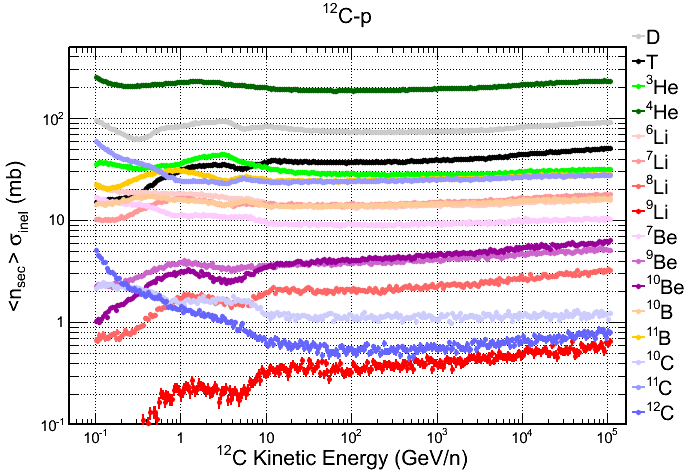}
\includegraphics[width=0.325\columnwidth,keepaspectratio,clip]{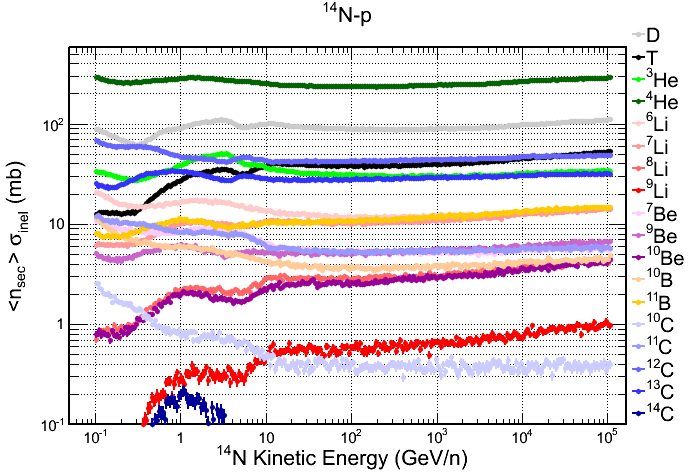}
\includegraphics[width=0.325\columnwidth,keepaspectratio,clip]{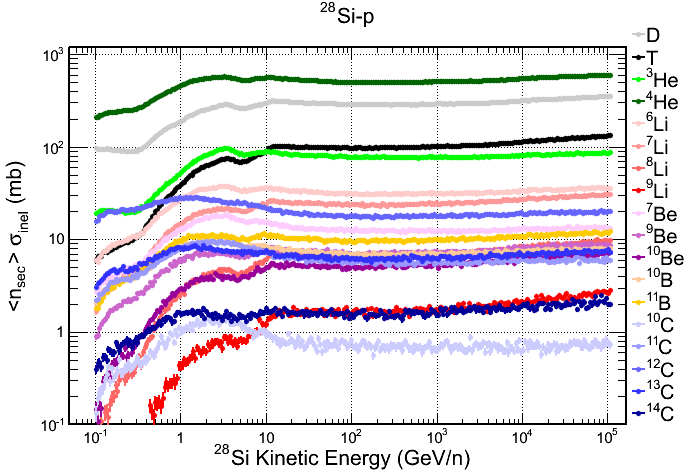}
\caption{Inclusive cross sections for the production of spallation nuclei in collisions 
of $^{12}$C, $^{14}$N and $^{28}$Si with p.
The plots show the cross sections for the production of Deuteron (gray markers), Triton (black markers) 
and the isotopes of He ($^{3}$He and $^{4}$He, green markers), 
Li ($^{6}$Li, $^{7}$Li, $^{8}$Li and $^{9}$Li, red markers), Be ($^{7}$Be and $^{9}$Be, magenta markers), 
B ($^{10}$B and $^{11}$B, orange markers) and 
C ($^{10}$C, $^{11}$C, $^{12}$C, $^{13}$C and $^{14}$C, blue markers). 
Lighter (darker) color shades correspond to lighter (heavier) isotopes.}
\label{FigSpallXsec}
\end{figure*}

With these results in hand, we can present the outcome of our analysis for all stable 
particles (including radionuclides whose decay is not taken into account at this stage).
In Fig.~\ref{FigpHeInclXsec} we show the total inclusive cross sections for the production of 
protons, electrons and positrons, gamma rays, electron neutrinos and antineutrinos,
muon neutrinos and antineutrinos and deuterons, as a function of the kinetic energy per nucleon of the projectile
in the p-p, p-$^{4}$He and $^{4}$He-$^{4}$He collisions.
In Fig.~\ref{FigSpallXsec} we show some relevant results among all the calculations we
performed for the heavy nuclei. We plot the inclusive cross sections of to the the interactions of $^{12}$C, $^{14}$N and $^{28}$Si 
with protons, yielding deuterons, tritons and all the isotopes of He, Li, Be, B and C.

We remark that this is the first time in which such a complete sample of nuclear cross sections are computed in a consistent way
with a single numerical code. Due to the completeness and accuracy of these computations, we point out that the dataset we 
produced can be used by the community working on CR physics to constrain the properties of CR transport in a more solid way, 
as we will show in the next Section.

\section{Application to the Galactic emission}

In order to evaluate the emission from the Galaxy we use a customized version of the propagation code
{\tt DRAGON} (2D version)\footnote{A version of {\tt DRAGON} code is available for download at the link http://www.dragonproject.org/.} 
in which we implemented the present parameterizations in the secondary production routine.

We study the spectra of the following species: protons, Helium, Boron, Carbon, 
electrons, and positrons; for each one we evaluate the contribution of secondaries originating from nuclear interactions.
In our simulation we assume that the ISM is composed of Hydrogen and Helium with relative abundances $1:0.1$.
The $\beta^{\pm}$ decays and the electron captures of unstable isotopes are taken into account according to their lifetimes. 
In the current version of the \texttt{DRAGON} code the daughter nucleus carries out all the energy of the parent nucleus, while the other products 
(i.e. leptons) are discarded.

The purpose of this section is to discuss the implications of the present parameterizations on some reference diffusion models.
The setup we are using is standard: CR transport properties are homogeneous and isotropic; the scalar diffusion coefficient depends on 
the particle rigidity $R$ and on the distance from the Galactic Plane $z$ according to the following parameterization:
$D = D_0 \, \beta^{\eta} \, \left({\frac{R}{R_0}}\right)^{\delta} \, e^{|z|/z_t}$
where
$D_0$ is the diffusion coefficient normalization at the reference rigidity $R_0 = 4\units{GV}$;
$\eta$ effectively describes the complicated physical effects that may 
play a major role at low energy (below $1 \units{GeV}$), e.g. the dissipation of Alfv\'en waves 
due to the resonant interaction with CRs (here $\beta$ represents as usual the particle velocity 
in units of the speed of light);
$\delta$ is the spectral index: it is constrained by the data on the 
light nuclei ratios, in particular by the measurements of the B/C ratio;
$z_t$ is the scale height of the diffusive halo of the Galaxy. 

Our starting point consists of two diffusion models 
considered as a reference in several previous works (see e.g. \cite{Evoli:2012ha}) labelled as KRA 
($\delta = 0.5$) and KOL ($\delta = 0.33$): they are mainly tuned on the PAMELA B/C data. 
We also present a new model tuned on the recent preliminary AMS02 
B/C data with $\delta = 0.42$ based on the one presented in \cite{Evoli:2015}.
 
In all cases the injected spectra up to Silicon nuclei are described by a broken power law with a single break at 10 GeV.
%
In all models the spectral indices of p, He and nuclei below the break are 2.05, 2.18 and 2.20 respectively.
In the KOL/KRA/AMS models the indices above the break of p, He and nuclei are 2.48/2.33/2.43, 2.40/2.24/2.32 and 2.60/2.40/2.50 respectively.
The source term distribution is taken from \cite{Ferriere:2001rg}. The gas density distribution is taken from the public {\tt Galprop} version \cite{galprop_website,galprop1998,galprop2002}.
We also consider a primary electron+positron extra component with harder injection spectrum in order to explain the anomalous rise of the positron fraction above $10 \units{GeV}$ reported by PAMELA, Fermi-LAT and AMS02. This behaviour cannot be reproduced in 
the standard scenario in which positrons only originate from spallation of protons, helium and heavier nuclei on interstellar gas. 

In order to compare our computations with the data, we have to consider that, in the final stage of their propagation, 
charged particles enter the sphere of influence of the Sun. Here they diffuse in the Heliospheric 
magnetic field, and suffer adiabatic energy losses and convection due to the presence of the solar wind: 
this process is called {\it solar modulation} and is relevant for low energy ($< 10 \units{GeV}$) particles. 
In this work we adopt the so-called \textit{``force field approximation''}~\cite{GA}, a simplified 
description in which a single free parameter is involved: the 
modulation potential $\phi$.

\begin{figure*}[!ht]
\begin{center}
\begin{tabular}{cc}
\includegraphics[width=0.45\columnwidth,keepaspectratio,clip]{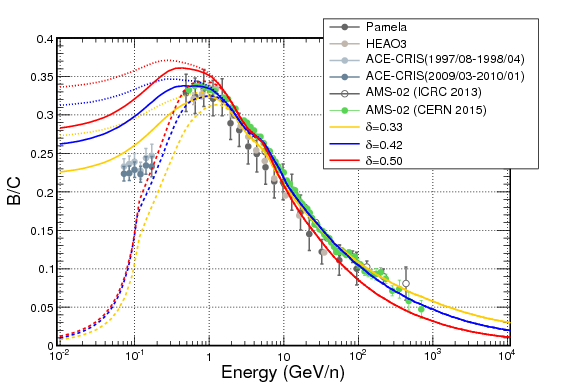} &
\includegraphics[width=0.45\columnwidth,keepaspectratio,clip]{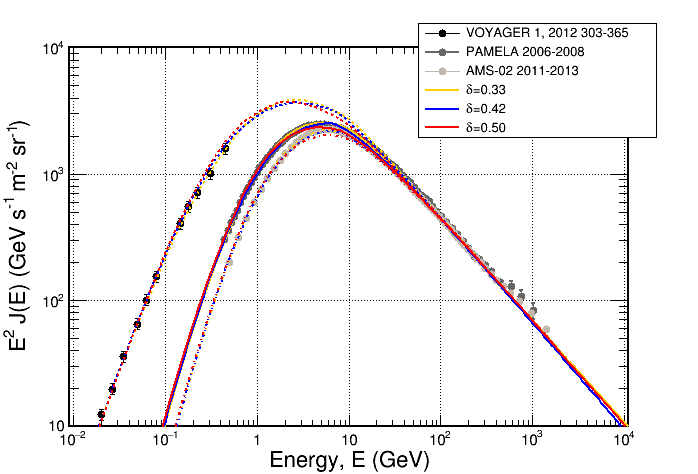} \\
\includegraphics[width=0.45\columnwidth,keepaspectratio,clip]{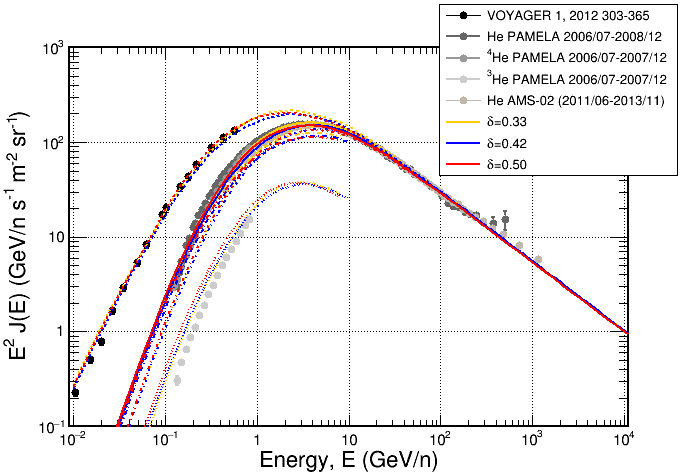} &
\includegraphics[width=0.45\columnwidth,keepaspectratio,clip]{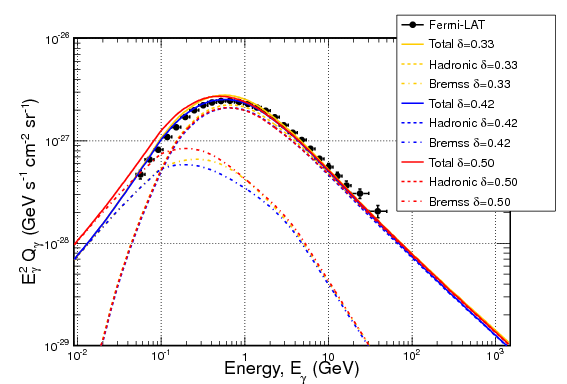}
\end{tabular}
\end{center}
\caption{DRAGON results. The models are: $\delta=0.33$ (KOL), $\delta=0.42$ and $\delta=0.50$ (KRA). 
Top left: Boron to Carbon ratio; bottom left: Helium; top right: proton, comparison with PAMELA, AMS02 and VOYAGER1~\cite{stone2013} data. Dashed line: unmodulated intensity; solid (dotted) line: modulated intensity by means of the force field approximation with $\phi$=0.42 (0.62) GV (some data point are quoted from~\cite{crdb}). The bottom right panel shows the differential gamma-ray emissivity as a function of the gamma-ray energy, 
evaluated by folding the proton and the Helium (hadronic), electron (bremsstrahlung) LIS with the $\gamma$-ray production cross sections (the data points are quoted from~\cite{jm2015}).}
\label{FigDragonFluka}
\end{figure*}

Given all these ingredients, our method is the following: we compare the current data to 
the computations performed with our modified version of {\tt DRAGON} with the new cross section parameterizations and we retune the model accordingly.
Our results are summarized in Fig.~\ref{FigDragonFluka}.
Very remarkably, we find that it is possible to reproduce correctly all the relevant observables (namely, proton, Helium, light nuclei spectra, and B/C ratio) with the reference models considered, 
after a rescaling of the diffusion coefficient normalization ($D_0$) and a minor fine-tuning of the other parameters.
For the KOL model we use $D_{0}$ = $4.25\times 10^{28}\units{cm^2~s^{-1}}$, $\eta = 0$ and $v_{\rm A} = 33\units{km~s^{-1}}$ (Alfv\'en velocity), while
for the KRA model we use $D_{0}$ = $2.8\times 10^{28}\units{cm^2~s^{-1}}$, $\eta = -0.4$ and $v_{\rm A} = 17.5\units{km~s^{-1}}$. 
Finally for the AMS02 model we use $D_{0}$ = $3.35\times 10^{28}\units{cm^2~s^{-1}}$, $\eta = -0.4$ and $v_{\rm A} = 17.5\units{km~s^{-1}}$. 

For the three models we use the solar modulation parameter $\phi$ = 0.42$\units{GV}$ and 0.62$\units{GV}$ for all the species to reproduce the PAMELA and AMS02 data respectively. 
The differences of these results with respect to the PAMELA and AMS02 proton data are within $\pm 10\%$ or better.
Clearly, the model with $\delta=0.42$ is designed to have the best compatibility with the AMS02 preliminary B/C data.
%
%
It is worth to point out that in these three models we do not consider the presence of the break around 
$300 \units{GeV}$ in the proton and nuclei spectra, since it is beyond the aim of this paper. However, with this complete sample of inclusive cross sections
we do not need to introduce the concept of nuclear enhancement factor to describe secondary production in interaction 
among heavier nuclei, extrapolating the results of p-p interaction~\cite{mori2009}. 
In particular, this factor depends on the shape of the spectrum (see for instance \cite{Kachelriess:2014mga}) and it does not work in case of abrupt change in the CR spectra (as in case of breaks).

Here we want to stress that with this new evaluation we have a much better insight on the nuclear processes, since 
the cross-sections can be computed accurately in a wide energy range. 
%
For example, given the more accurate determination of the gamma-ray emissivity, we can turn our attention
to the Local Interstellar Spectra (LIS) of protons and Helium nuclei, not affected by solar modulation.
This quantity can be inferred from the Fermi-LAT gamma-ray data~\cite{jm2015} and its determination is 
useful in order to constrain the propagation models described above, since each of them implies a different shape and normalization for the LIS. 
The bottom-right panel of Fig.~\ref{FigDragonFluka} shows the gamma-ray emissivity calculated by means of the proton, Helium and electron LIS.
We calculate the contribution associated with the hadronic collisions and bremsstrahlung by folding the proton, Helium and electron LIS 
with the corresponding $\gamma$-ray production cross sections on the ISM with relative abundances of H : He : C : N : O : Ne : Mg : Si =
$1:0.096:4.65\times 10^{-4}:8.3\times 10^{-5}:8.3\times 10^{-4}:1.3\times 10^{-4}:3.9\times 10^{-5}:3.69\times 10^{-5}$.
An inspection of this plot shows that at low energies the contribution from electron bremsstrahlung is not negligible. 
This contribution mainly depends on the electron spectrum, that should also be constrained taking synchrotron radiation into account (see for instance \cite{DiBernardo:2012zu,Orlando:2013ysa}).

\section{Conclusions}

We have used the {\tt FLUKA} simulation code to evaluate the inclusive cross sections for the production of
stable secondary particles and as well as nuclei from spallation in the interactions of several cosmic-ray projectiles 
with different target nuclei.
%

As an example of application we implemented the values of the cross sections in a custom version of the
{\tt DRAGON} CR propagation code, showing that it is possible to reproduce with accuracy the measured 
CR spectra and light nuclei ratios as well as the gamma-ray emissivity.

\end{document}